\documentclass[12pt,preprint]{aastex}

\usepackage{epstopdf}

\def\gs{\mathrel{\raise0.35ex\hbox{$\scriptstyle >$}\kern-0.6em
\lower0.40ex\hbox{{$\scriptstyle \sim$}}}}
\def\ls{\mathrel{\raise0.35ex\hbox{$\scriptstyle <$}\kern-0.6em
\lower0.40ex\hbox{{$\scriptstyle \sim$}}}}

\def\spose#1{\hbox to 0pt{#1\hss}}
\def\simlt{\mathrel{\spose{\lower 3pt\hbox{$\mathchar"218$}}
     \raise 2.0pt\hbox{$\mathchar"13C$}}}
\def\simgt{\mathrel{\spose{\lower 3pt\hbox{$\mathchar"218$}}
     \raise 2.0pt\hbox{$\mathchar"13E$}}}

\newcommand{\um}{\,$\mu$m\,}

\newcommand{\spitz}{{\sl Spitzer}}

\newcommand{\msun}{\,$\rm{M}_{\odot}$}

\shorttitle{$z$\,$\sim$\,2 radio-loud sources with deep 9.7\um\ absorption}
\shortauthors{Sajina et al.}

\received{2007 July 6}
\begin{document}

\title{Discovery of Radio Jets in $z$\,$\sim$\,2 ULIRGs with Deep 9.7\um\ Silicate Absorption}

\author{Anna Sajina\altaffilmark{1}, Lin Yan\altaffilmark{1}, Mark Lacy\altaffilmark{1}, Minh Huynh\altaffilmark{1}}

\altaffiltext{1}{\spitz\ Science Center, California Institute of Technology, Pasadena, CA 91125}

\begin{abstract}
Recent Spitzer observations have revealed a substantial population of $z$\,$\sim$\,2 ULIRGs with deep silicate absorption ($\tau_{9.7}$\,$>$\,1). This paper reports a 20cm radio study of such a sample to elucidate their physical nature. We discover that a substantial fraction (40\%) of deep silicate absorption ULIRGs at $z$\,$\sim$\,2 are moderately radio-loud with $L_{1.4\rm{GHz}}$\,=\,$10^{25}$\,--\,$10^{26}$\,WHz$^{-1}$. This is in strong contrast with $z$\,$\ls$\,1 radio galaxies and radio-loud quasars where none of the sources with available IRS spectra have $\tau_{9.7}$\,$>$\,1.  In addition, we observe radio jets in two of our sources where one has a double lobe structure $\sim$\,200\,kpc in extent, and another shows a one-sided jet extending $\sim$\,90\,kpc from the nucleus. The likely high  inclination of the latter, coupled with its deep silicate absorption, implies the mid-IR obscuration does not share an axis with the radio jets. 
These sources are highly obscured quasars, observed in the transition stage after the birth of the radio source, but before feedback effects dispel the ISM and halt the black hole accretion and starburst activity.  

\end{abstract}


\section{Introduction}
The local relationship between the bulge mass and black hole mass \citep{magorrian98} suggests that the stellar mass build-up and black hole growth were co-eval. This is supported by the number density evolution of quasars and radio galaxies \citep{wall05,richards06} being similar to the evolution of the star-formation rate (SFR) density with redshift \citep[e.g.][]{sfrh_goods}.  Observations of significant far-IR emission potentially suggest starbursts associated with high-$z$ quasars, and radio galaxies \citep[e.g.][]{willott02,dusty_radio,beelen06} and indicate rapid growth of the host galaxy likely fueled by mergers \citep{hutchings06,role_mergers}. Studies of the young stellar populations of radio galaxies also suggest merger-driven starbursts \citep{tadhunter05}.  Observations of $z$\,$\sim$\,2 radio galaxies indicate the powerful outflows that will disperse the surrounding ISM \citep{outflows_rl}.

To study the process of a quasar emerging from its dusty cocoon, we need transition Ultraluminous Infrared Galaxies (ULIRGs) -- i.e. containing a still dust enshrouded quasar \citep[e.g.][]{cs01}.  Unlike infrared, optical, or X-ray diagnostics, extreme radio luminosity is a dust-independent, unambiguous indicator of a highly obscured AGN \citep[e.g.][]{alejo05,donley05}. 

We have a sample of {\sl Spitzer}-selected $z$\,$\sim$\,2  ULIRGs whose continuum-dominated mid-IR spectra already imply the presence of an AGN \citep{yan07,sajina07}. In this letter, we highlight the seven sources of this sample which are radio-loud, confirming the AGN, while their deep 9.7\um\ silicate absorption indicates significant obscuration.  

Throughout this letter we adopt the $\Lambda$CDM cosmology with $\Omega_{\rm{M}}$\,=\,0.3, $\Omega_{\Lambda}$\,=\,0.7, and $H_0$\,=\,70\,$\rm{km}\rm{s}^{-1}\rm{Mpc}^{-1}$.

\section{The sample \label{sec_sample}}
Our sources are part of a sample of 52 high-$z$ ULIRGs with IRS spectra \citep{yan07}.  In summary, the sample is selected from the {\sl Spitzer} First Look Survey (FLS). They are selected for having 24\um\ flux density brighter than 0.9mJy, and red 24-to-8\um\ and 24-to-R colors\footnote{(1) $R(24,8) \equiv \log_{10}(\nu f_{\nu}(24\mu m)/\nu f_{\nu}(8\mu m) \simgt 0.5$; (2) $R(24,0.64) \equiv \log_{10}(\nu f_{\nu}(24\mu m)/\nu f_{\nu}(0.64\mu m) \simgt 1.0$.}.  Of these, 47 have IRS-based redshifts, with the majority (33/47\,=\,72\%) at 1.5\,$<$\,$z$\,$<$3.2. The analysis of the IRS spectra, and derivation of PAH luminosities, equivalent widths, and silicate feature optical depths\footnote{Our 9.7\um\ optical depth is measured from an extinction-corrected continuum and relates to the observed depth of the silicate feature as $\tau_{9.7}$\,=\,1.4\,$\tau_{Si}$ \citep[see][for details]{sajina07}.} ($\tau_{9.7}$) is presented in \citet{sajina07}.  All sources in our sample show evidence of both star-formation and AGN activity; however, based on their PAH-strengths vs. mid-IR continua, the `weak-PAH' (EW$_{7.7}$\,$<$\,0.8\um) sources are believed to be energetically dominated by their AGN.  About 60\% of the these weak-PAH sources have significant 9.7\um\ silicate absorption with $\tau_{9.7}$\,$>$\,1.  Roughly half of these show extreme  optical depths ($\tau_{9.7}$\,$>$\,3). At such high obscurations the IR spectra of AGN and compact, nuclear starbursts are indistinguishable \citep{roussel06,levenson07}.  Of the total sample, 60\% are detected in the VLA 20\,cm map of the FLS \citep{condon03}, and 48\% are detected  in the GMRT 610MHz map of the field \citep{garn07}. For a full description of the radio data see Sajina et al. (2007, in prep.). 

In this letter, we only look at the 28 $z$\,$>$\,1.5 continuum-dominated (i.e. likely AGN dominated) sources.  We use their radio emission to reveal the hidden AGN.  We take  $L_{1.4\rm{GHz}}$\,$>$\,$10^{25}$\,WHz$^{-1}$ as our `radio-loud' criterion \citep[e.g.][]{def_rl}, while we discuss the more standard radio-to-optical ratio criterion in \S\,\ref{sec_mir}.  The seven sources which satisfy this criterion are listed in Table~\ref{table_sample} along with a summary of their mid-IR properties from our earlier papers \citep{yan07,sajina07} as well their radio fluxes and luminosities from Sajina et al. (2007, in prep).   These seven sources represent 25\% of the $z$\,$\sim$\,2 continuum-dominated sources. For comparison, optically-selected quasars have $\sim$\,10\,-\,20\% radio-loud fraction \citep{def_rl}.  

\section{Analysis}
\subsection{Mid-IR obscuration \label{sec_tau}}
\begin{figure}[!ht]
\begin{center}
\plotone{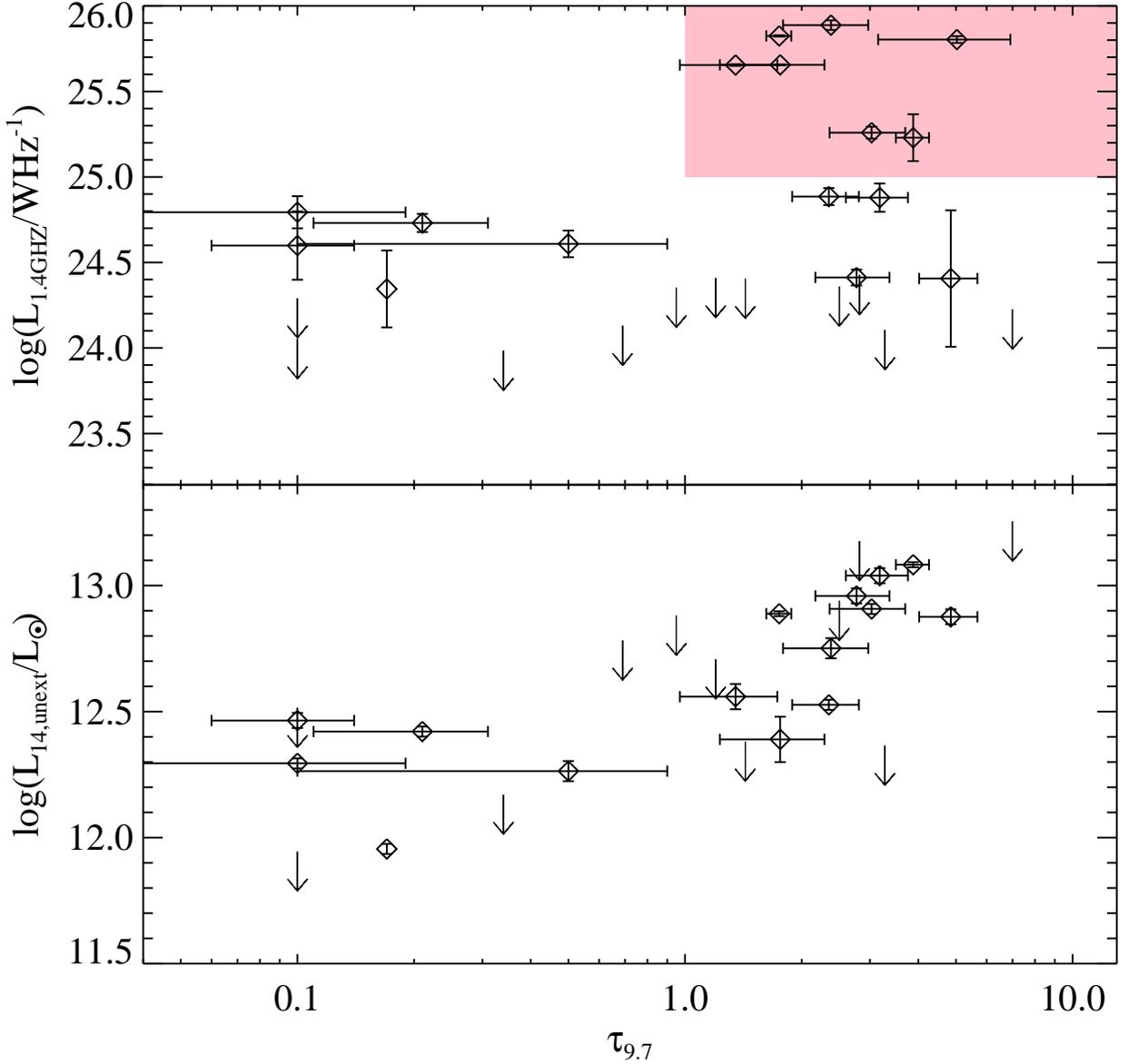}
\end{center}
\caption{{\it Top:} The 1.4\,GHz luminosity vs. the 9.7\um\ optical depth. For our sample, only sources with strong silicate features ($\tau_{9.7}$\,$>$\,1) meet the radio-loud criterion ($L_{1.4\rm{GHz}}$\,$>$\,$10^{25}$WHz$^{-1}$).  The pink box highlights the region occupied by our seven radio-loud sources.  {\it Bottom:} Mid-IR continuum luminosity vs. $\tau_{9.7}$. After extinction correction, $L_{14\mu\rm{m}}$ of the high-$\tau$ sources are higher than those of the low-$\tau$ sources, possibly providing clues as to why their radio luminosities are higher as well. \label{radio_tau}}
\end{figure}

The top panel of Figure~\ref{radio_tau} shows that all of our radio-loud sources have deep silicate absorption features ($\tau_{9.7}$\,$\sim$\,1\,--\,6). Among our $z$\,$\sim$\,2 high-$\tau$ sources, $\sim$\,40\% are radio-loud, making this an exceptionally efficient means of selecting high-$z$ radio-loud sources.  By contrast, none of the low-$\tau$ sources are radio-loud. The observed $L_{14\mu\rm{m}}$ of the low- and high-$\tau$ sources are not significantly different (this holds for $L_{\rm{FIR}}$ as well; Sajina et al. 2007 in prep.). However, the extinction corrected 14\um\ continua are significantly higher for the deeply absorbed sources (bottom panel in Figure~\ref{radio_tau}).  Higher mid-IR luminosities might indicate any of greater black hole masses, greater accretion rates, or more compact dust geometry leading to hotter dust. At present it is unclear which, if any, of these reasons might explain the lack of radio-loud sources among our low-$\tau$ continuum-dominated sources.  \citet{shi06} show a correlation between the gas column densities ($N_{\rm{H}}$) and $\tau_{9.7}$. Extrapolating this to the optical depths of our sources, we find $N_{\rm{H}}$\,$\gs$\,$10^{27}$\,cm$^{-2}$. If confirmed, these sources qualify as Compton-thick quasars. It is however likely that this trend breaks down due to differences in the properties and geometry of the obscuring medium (for example a more uniform, screen-like obscuration as opposed to a more clumpy one).  Initial {\sl Chandra} observations support strong X-ray obscuration (Bauer et al. 2008 in prep) given non-detections in 30\,ks exposures. Lastly, five of the seven meet the \citet{alejo05} Compton-thick quasar criteria (MIPS15880 has too much 3.6\um\ flux, while MIPS16122 has too much 1.4\,GHz flux).  

\subsection{Spectral indices \label{sec_alpha}}
The radio spectral indices\footnote{Defined as $\alpha$\,=$\log(S_{610\rm{MHz}}/S_{1.4\rm{GHz}})/\log(610\rm{MHz}/1.4\rm{GHz})$.} ($\alpha$) based on the integrated fluxes at 610\,MHz and 1.4GHz are listed in Table~\ref{table_sample}.  The different resolution of the two images means that the 1.4\,GHz map might be missing extended flux relative to the 610\,MHz.  We estimate the strength of this effect by comparing the $\alpha$ values based on the integrated or peak 610\,MHz fluxes. We find that our $\alpha$'s can be overestimated by up to $\sim$\,0.2\,dex. This mean up to 0.1\,dex overestimation of the radio luminosities, which is not significant for our conclusions. The largest effect is found for our strongest radio source, MIPS16122, where $\alpha$ ranges between -0.0 and -0.5 for the peak and integrated flux values respectively.  This is our only potentially flat-spectrum source. The typically steep spectra we find are consistent with the $\alpha$\,$\sim$\,-1 of $z$\,$\sim$\,2 Type-2 quasars found by \citet{alejo06}. Such steep slopes are associated with evolved (i.e. lobe-dominated) radio sources \citep{alejo06}, while flat spectra are generally core-dominated sources.  Given our radio resolution, and the redshifts of these sources, we are only sensitive to extended structures of $\gs$\,100\,kpc.  Below we discuss the two sources with observed radio jets. 

\subsection{Extended radio emission \label{sec_frii}}

\begin{figure}[!ht]
\begin{center}
\plotone{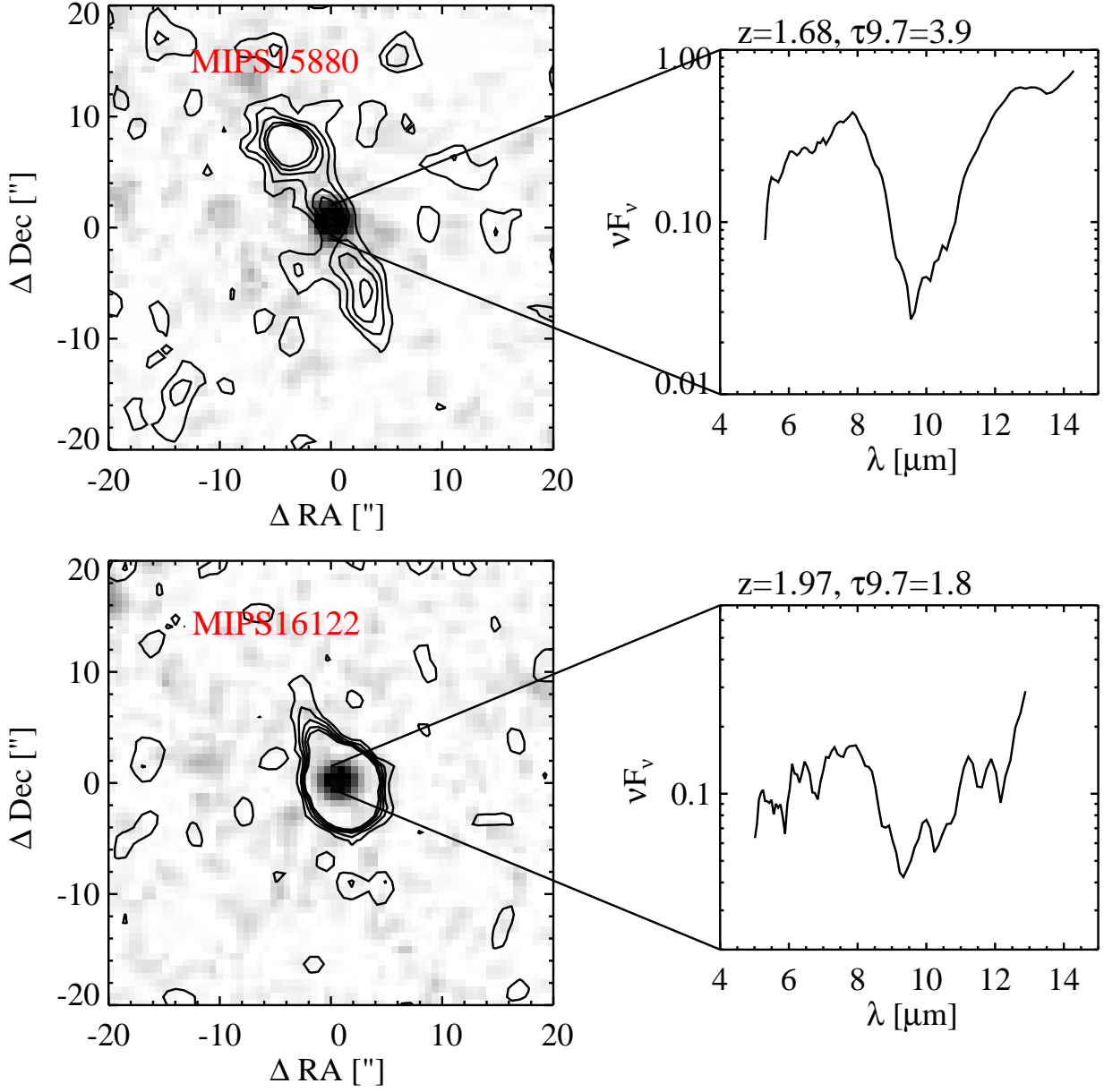}
\end{center}
\caption{The VLA 1.4\,GHz contours (1\,--\,5\,$\sigma$) overlaid on the MIPS24\um\ image for MIPS15880 ({\it top}), and MIPS16122 ({\it bottom}) respectively.  At the redshifts of both sources the radio beam ($\sim$\,5\arcsec) subtends $\approx$42\,kpc. None of the extended structures have counterparts in the $R$-band or IRAC3.6\um\ images. On the right hand side are shown the IRS mid-IR spectra of both sources. In both cases the redshifts are derived from the silicate feature. \label{example_rl}}
\end{figure}

In \S\,\ref{sec_alpha}, we discussed that our spectral indices suggest evolved radio sources. MIPS15880 shows a transition FRI/FRII structure with a diameter of $\sim$\,200\,kpc (see Figure~\ref{example_rl}), which is near the mean size for double-lobed structures \citep{devries06}.  Given that such moderate luminosity radio sources have hot spot advance speeds in the range $\sim$\,0.01\,--\,$0.1c$, this size implies a radio source age of $\sim$\,$10^7$yrs \citep[e.g.][]{willott02}. 

The  bottom panel of Figure~\ref{example_rl} (MIPS16122) shows the characteristic tadpole shape indicative of a one-sided jet, suggesting that the jet axis is close to the line-of-sight ($<<$\,45$^{\circ}$). It spans $\sim$\,90$/sin(i)$\,kpc, where $i$ is the inclination angle.  Higher resolution radio data is, however, required to confirm the morphology of the radio source. We see no counterpart to the extended emission in our {\sl HST} NICMOS image \citep{dasyra07}.

\subsection{Radio vs. mid-IR emission \label{sec_mir}}
\begin{figure}[!ht]
\begin{center}
\plotone{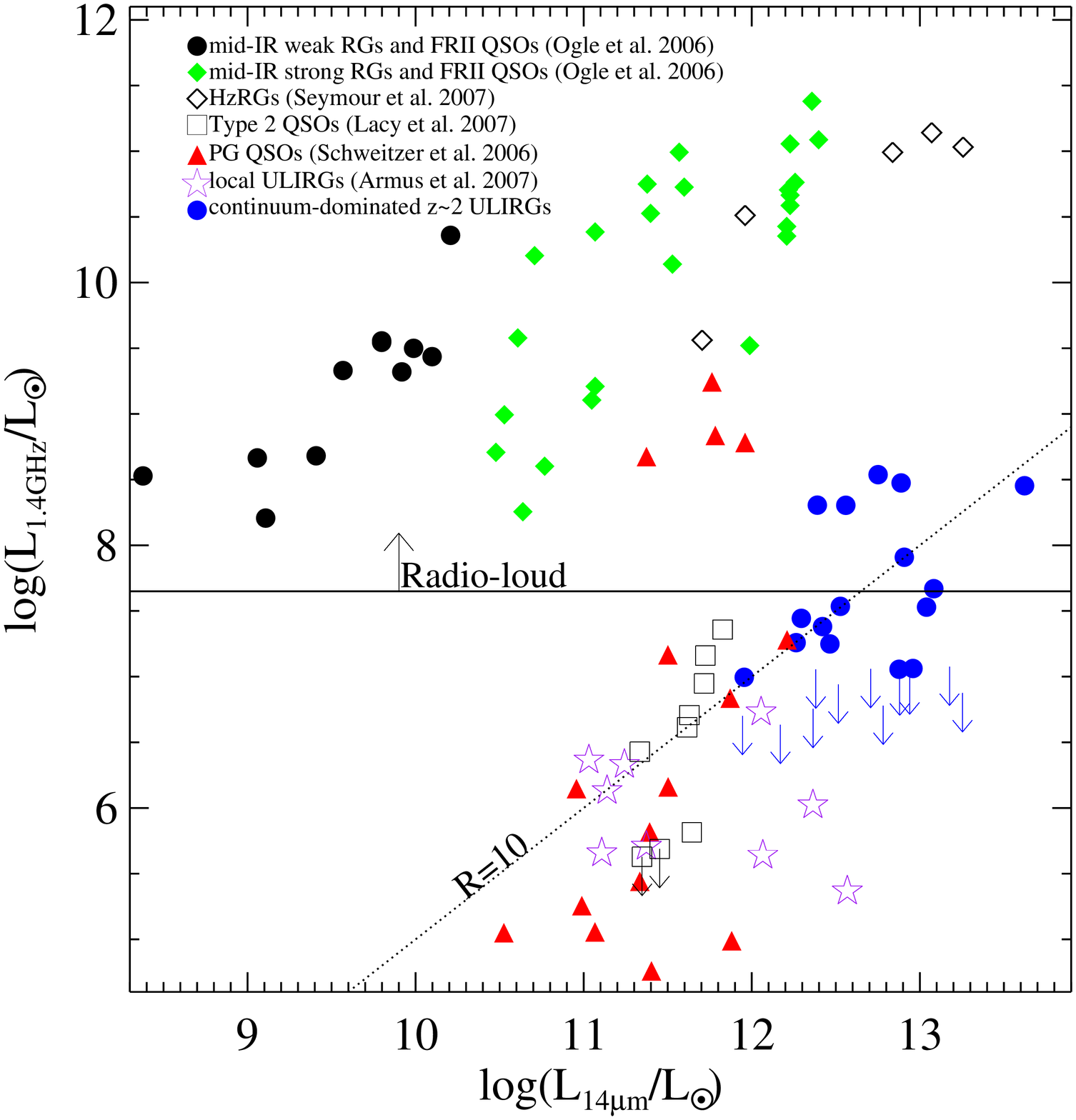}
\end{center}
\caption{A comparison between our sources and other radio-loud or ULIRG populations.  \citet{ogle06}quote $L_{15}$, which is close enough to $L_{14}$ for our purposes. The Type 2 quasars we plot are all at $z$\,$\sim$\,0.4\,--\,0.8 so that the observed 24\um\ flux is close to the rest-frame 14\um. The HzRGs shows are those with both 24\um\ and 70\um\ detections \citep{seymour07}, we interpolated between these fluxes to derive the rest-rame 14\um. Both for our sample and the ULIRGs we use extinction corrected $L_{14\mu\rm{m}}$. The 1.4\,GHz fluxes of the PG quasars, Type 2 quasars, and ULIRGs are taken from NED. The 1.4\,GHz is derived by assuming $\alpha$\,=\,-0.7.  The dotted line represents the classic radio-to-optical ratio definition of `radio-loud' (see text for details).  \label{radio_tau_all}}
\end{figure}

Figure~\ref{radio_tau_all} puts our sources in context with related populations.  A number of assumptions were necessary to put the various disparate samples on the same plane (see caption Figure~\ref{radio_tau_all}). These assumptions are crude, but acceptable given the dynamic range of this figure. 

Does our classification agree with the more common `radio-loud'  definition, $R$\,$\equiv$\,f5GHz/f4400\AA\,$>$\,10? Given the significant obscuration of our sources this definition is not immediately applicable. We translate f4400\AA\ to f14\um\ luminosities based on the average quasar spectrum \citep{richards06_sed}, and assume $\alpha$\,=\,-0.7 in the radio, which leads to an equivalent criterion: $R_{\rm{IR}}$\,$\equiv$\,$L_{1.4\rm{GHz}}$/$L_{14\mu\rm{m}}$\,$>$\,$10^{-5}$ (dotted line in Figure~\ref{radio_tau_all}).  Four of our sources remain `radio-loud' in this definition as well, while three are borderline, although possibly still radio-loud given the rough derivation above.  In addition, the screen extinction correction we apply to $L_{14\mu\rm{m}}$ is likely to overestimate the mid-IR luminosities. With no strong dichotomy between our `radio-loud' and `radio-quiet' sources, are we not subject to rather arbitrary definitions? Obviously, we see extended emission among the radio-loud sources and not the rest of the sample.  Only four of the non-radio-loud sources have detections at 610\,MHz, implying a generally flatter radio spectrum, than for the radio-loud sources.  Lastly, the surprising difference between high-$\tau$ and low-$\tau$ sources (see \S\,3.1).  These support the view of our radio-loud sources as a distinct population.  
 
\subsection{Black hole masses} 
The total IR luminosities ($\sim$\,3\,--\,4\,$L_{14\mu\rm{m}}$; Sajina et al. 2007 in prep.) are likely $\approx$\,$L_{\rm{bol}}$ for such highly obscured sources.  Based on these, and assuming Eddington accretion, we estimate black hole masses of $\sim$\,2\,--\,6\,$\times$\,$10^8$\msun. This is supported by the empirical relation between radio luminosity, $M_{\rm{BH}}$ and accretion rate \citep{lacy01}, which leads to $\langle M_{\rm{BH}}\rangle$\,$\sim$\,3.0\,$\times$\,$10^8$\,M$_{\odot}$ again assuming Eddington accretion.  
  
Our sources do not represent a complete census of massive black holes at $z$\,$\sim$\,2, and are likely to experience further growth between then and $z$\,$\sim$\,0. With this in mind, can we reconcile such massive black holes at $z$\,$\sim$\,2 with the space density of massive black holes today? The density of our $z$\,$\sim$\,2 heavily obscured radio-loud sources is $\sim$\,3\,$\times$\,$10^{-7}$\,Mpc$^{-3}$.  \citet{lauer07} estimate the space density of black holes with $M_{\rm{BH}}$\,$\ge$\,3\,$\times$\,$10^8$\msun\ (typical of our sources) is $\sim$\,3\,$\times$\,$10^{-4}$\,Mpc$^{-3}$ locally.  
This generous gap between the space densities implies we are not overproducing black holes at $z$\,$\sim$\,2.

\section{Discussion}
Our study for the first time directly demonstrates the existence of deep silicate absorption ($\tau_{9.7}$\,$\sim$\,1\,--\,6) radio-loud sources at $z$\,$\sim$\,2. This is in strong contrast with $z$\,$\ls$\,1 radio-loud populations where none have $\tau_{9.7}$\,$>$\,1 \citep{haas05,ogle06}. The high obscurations of our sources is largely a selection bias, as our sample was selected to be mid-IR bright and red (\S\,2).  However, {\it high radio fluxes are not part of our selection}, and hence the 40\% radio-loud fraction of our high-$\tau$ $z$\,$\sim$\,2 sources is a real effect. This implies a direct or indirect coupling between the radio source and surrounding ISM.  \citet{weedman06} present a $z$\,$\sim$\,2 sample similar to ours, but including a radio selection. A quick analysis of their sample yields two sources that are both radio-loud, and have $\tau_{9.7}$\,$>$\,1, further supporting to our results.   

The much greater $L_{14\mu\rm{m}}/L_{1.4\rm{GHz}}$ ratios of our sources relative to local radio-loud sources imply we are observing these sources in the brief window between the birth of the radio source and before feedback effects have shed the dusty envelope and halted the starburst and/or black hole accretion. \citet{willott02} argue that the IR-brightness is anti-correlated with the age of the radio source consistent with a radio source-driven feedback scenario. Given the age of MIPS15880 ($\gs$\,$10^7$\,yrs), its radio jets are not necessarily powerful enough to accomplish this.  But this may be a function of radio luminosity. MRC1138-262, a $z$\,=\,2.2 radio galaxy more than two orders of magnitude more radio-luminous than our sources, does show powerful outflows, which are probably driven by lateral shocks from the radio jets \citep{outflows_rl}.  A mid-IR study of a sample of $z$\,$\sim$\,2 radio galaxies, covering a range of radio luminosities, would therefore provide very useful constraints on radio jet/ISM interactions at the epoch when radio galaxies are in the process of shedding their dusty envelopes. 

The above does not explain why our sources should have deeper silicate absorption than local radio-loud populations. The answer hinges on the unknown origin of the mid-IR obscuration -- s.a. dusty torus or host galaxy. The deep silicate features of our sources  imply optically thick obscuring media with high filling factors (i.e. likely puffed-up geometry, but not 4\,$\pi$ given the radio jets) as they have to effectively `hide' the hot inner regions of the AGN \citep{pk92,levenson07}. AGN tori are believed to be clumpy, which precludes high filling factors \citep{elitzur06}, consistent with their observed weak silicate features \citep{shi06}. Higher $\tau_{9.7}$ can be due to a cold dust screen \citep{ogle06}, such as expected from a dusty host galaxy. High-$z$ radio galaxies/radio-loud quasars have been detected in the sub-mm indicating cold dust \citep{willott02,dusty_radio}. We already know that the number density of deeply obscured sources is about 10 times higher at $z$\,$\sim$\,2 than at $z$\,$\sim$\,0 \citep{sajina07}. Lastly, MIPS16122 is consistent with a jet axis close to the line-of-sight ($<<$\,45$^{\circ}$).  The dusty torus of an AGN is believed to be coupled with the accretion disk \citep[e.g.][]{elitzur06}, implying a nearly face-on torus for MIPS16122, which is unlikely given its deep silicate absorption feature.  On the other hand, a mis-match between the orientation of the host galaxy and AGN is frequently observed \citep{schmitt02}. Thus, although the origin of the mid-IR obscuration remains an open question, host galaxy obscuration is a good candidate.  \\
{\it Acknowledgements:} Many thanks to Patrick Ogle for a critical reading of the manuscript prior to submission. We are also grateful to N. Seymour, K. Dasyra, B. Partridge, D. Lutz, and D. Scott for useful discussions. We are grateful to the anonymous referee for their helpfull comments. This work is based on observations made with the {\sl Spitzer} Space Telescope, which is operated by the Jet Propulsion Laboratory, California Institute of Technology under contract with NASA. Support for this work was provided by NASA through an award issued by JPL/Caltech. This research has made use of the NASA/IPAC Extragalactic Database (NED) which is operated by the Jet Propulsion Laboratory, California Institute of Technology, under contract with the National Aeronautics and Space Administration.

\clearpage

\begin{deluxetable}{ccccccccc}
\tablecaption{Summary of radio-loud sample properties\label{table_sample}}
\tablewidth{0pt}
\tablehead{\colhead{MIPSID} &  \colhead{$z$}  & \colhead{$\tau_{9.7}$} &\colhead{log($L_{\rm{PAH},7.7}$)} & \colhead{log($L_{14\mu\rm{m}}$)\tablenotemark{a}} & \colhead{$S_{1.4\rm{GHz}}$} & \colhead{$S_{610\rm{MHz}}$} & \colhead{$\alpha$} & \colhead{log($L_{1.4\rm{GHz}}$)} \\
 \colhead{} & \colhead{} & \colhead{} & \colhead{$L_{\odot}$} & \colhead{$L_{\odot}$} & \colhead{mJy} & \colhead{mJy} & \colhead{} & \colhead{WHz$^{-1}$} 
 }

\startdata

8327 & 2.48 & 2.4 & 10.62\,$\pm$\,0.21 & 12.40\,$\pm$\,0.04 & 1.40\,$\pm$\,0.06 & 3.44\,$\pm$\,0.14 &  -1.1 &  25.89\,$\pm$\,0.03\tablenotemark{b}  \\
15880\tablenotemark{c} & 1.68 & 3.9 & 10.60\,$\pm$\,0.09 & 12.51\,$\pm$\,0.01 & 0.60\,$\pm$\,0.03 & 1.35\,$\pm$\,0.06 & -1.0 & 25.02\,$\pm$\,0.14 \\
16059\tablenotemark{d} & 2.43 & 2.7 & 10.76\,$\pm$\,0.09 & 12.46\,$\pm$\,0.02 & 0.57\,$\pm$\,0.03 & --- &  --- &  25.26\,$\pm$\,0.04  \\
16122 & 1.97 & 1.8 & 10.41\,$\pm$\,0.21 & 12.13\,$\pm$\,0.09 & 2.82\,$\pm$\,0.14 & 4.34\,$\pm$\,0.12 & -0.5 & 25.66\,$\pm$\,0.003  \\
22204 & 2.08 & 1.6 & 11.01\,$\pm$\,0.10 & 12.63\,$\pm$\,0.01 & 1.89\,$\pm$\,0.08 & 4.67\,$\pm$\,0.10 & -1.1  & 25.82\,$\pm$\,0.004 \\
22558 & 3.2 & $>$\,5.6 & 10.96\,$\pm$\,0.12 & 12.88\,$\pm$\,0.01 & 0.62\,$\pm$\,0.03 & 1.55\,$\pm$\,0.12 & -1.1 & 25.80\,$\pm$\,0.02  \\
22277 & 1.77 & 1.4 & 10.44\,$\pm$\,0.19 & 12.36\,$\pm$\,0.05 & 1.57\,$\pm$\,0.07 & 4.79\,$\pm$\,0.11 & -1.3 & 25.65\,$\pm$\,0.005  
\enddata

\tablenotetext{a}{This is monochromatic $\nu L_{\nu}$. }

\tablenotetext{b}{This error accounts for the flux error alone,  $\sim$\,0.2\,dex error in $\alpha$ leads to an additional $\sim$\,0.1\,dex uncertainty.}

\tablenotetext{c}{The radio fluxes for MIPS15880 are the sum of the three components listed in the respective catalogues, we have checked that there is no significant extended flux missed. }

\tablenotetext{d}{MIPS16059 is undetected at 610\,MHz, since it falls in the noisy edge of the map. We assume $\alpha$\,=\,-0.7 for this source. }

\end{deluxetable}

\clearpage

\end{document}